\documentclass[%
 reprint,
superscriptaddress,
amsmath,amssymb,
aps,
pra
]{revtex4-1}

\usepackage{graphicx}
\usepackage{dcolumn}
\usepackage{bm}
\usepackage[colorlinks,linkcolor=blue,citecolor=blue,urlcolor=blue]{hyperref}
\usepackage{tikz}
\usepackage{subfigure}

\begin{document}

\preprint{APS/123-QED}

\title{
Topology of quadrupolar Berry phase of a Qutrit
}

\author{Rajeev Singh}
\affiliation{Department of Physics, Indian Institute of Technology (Banaras Hindu University), Varanasi, Uttar Pradesh 221005, India}
\author{Navneet Kumar Karn}
\affiliation{Department of Physical Sciences, IISER Kolkata, Mohanpur, West Bengal 741246, India}
\affiliation{CSIR-National Physical Laboratory
Dr. K.S. Krishnan Marg
New Delhi – 110012, India}
\author{Rahul Bhowmick}
\affiliation{Department of Physical Sciences, IISER Kolkata, Mohanpur, West Bengal 741246, India}
\author{Sourin Das}
\affiliation{Department of Physical Sciences, IISER Kolkata, Mohanpur, West Bengal 741246, India}

\date{\today}

\begin{abstract}
We examine Berry phase pertaining to purely quadrupolar state ($\langle \psi | \vec{S} | \psi \rangle = 0$) of a spin-$1$ system. 
Using the Majorana stellar representation of these states,  we provide a visualization for the topological (zero or $\pi$) nature of such quadrupolar Berry phase. We demonstrates that the $\pi$ Berry phase of quadrupolar state is induced by the Majorana stars collectively tracing out a closed path (a great circle) by exchanging their respective positions on the Bloch sphere. We also analyse the problem from the perspective of dynamics where a state from the quadrupolar subspace is subjected to a static magnetic field. We show that time evolution generated by such Hamiltonian restricts the states to the quadrupolar subspace itself thereby producing a geometric phase (of the Aharonov-Anandan type) quantized to zero or $\pi$. A global unitary transformation which maps the quadrupolar subspace to the subspace of purely real states proves a natural way of understanding the topological character of this subspace and its connection to the anti-unitary symmetries.
\end{abstract}

\maketitle

\section{\label{sec:level1}Introduction}
In recent times,  geometric phase has played a pivotal role in  our understanding of the physics of topological insulators where the topological properties of these band insulators could be understood in terms of the Berry curvature associated with band structure of these materials \cite{Kane2010,Zhang2011}.
This encompasses avatars of the geometric phase which {\it{can not}} be understood in terms of closed loop adiabatic evolution of the magnetic field applied to a spin-$S$ particle ($S=1/2,1,3/2...$), which is the conventional example of Berry phase\cite{Berry1984}. This has been a topic of much discussion in recent times triggered by the discovery of higher order topological insulators\cite{Benalcazar2017} which involves Hamiltonians having band structure supporting nonzero electric quadrupole and octupole (in general multi-poles)  moments  resulting in non-trivial Berry curvature. Effects of magnetic quadrupolar Berry phase in interacting many-body systems such as models of spin chain has also been studied\cite{Pollmann2014}.
More recent developments in fragile topological insulators\cite{fragile} or Euler insulators\cite{Ezawa} also has an interesting connection to qutrit system with quantized Berry phase\cite{Euler}. Motivated by these recent developments, we revisit this problem in a minimal setting of a spin $S=1$ system (or, equivalently a three level system) where such quadrupolar Berry phases can arise.

In 1994, Robbins and Berry\cite{Robbins_1994} showed that reversing the direction of externally applied magnetic field (say along the $z$-direction) acting on a spin-$S$ (where $S \in \mathbb{Z}$) results in a geometric phase factor of  $(-1)^{S}$ for the eigenstate of $\hat{S_z}$ operator with zero eigenvalue. 
Such a geometric phase may arise as a result of adiabatic rotation of the applied magnetic field such that it completes half cycle (one way journey between two antipodal points assuming the tip of the magnetic field to lay on a  sphere) and not a full cycle. Note that, even though the eigenstate of $\hat{S_z}$ operator with zero eigenvalue  does not have a direct coupling to the external magnetic field,  it nevertheless responds to its adiabatic evolution via the appearance of this geometric phase. It can be understood as follows -  rotating the magnetic field by half cycle reorganizes all states in the Hilbert space except this one. This reorganization of  states leads to the geometric phase factor of  $(-1)^{S}$. For $S=1$, the phase factor is $-1$, which corresponds to $\pi$ Berry phase.  
For $S=1$, the condition  $\langle \psi | \vec{S} | \psi \rangle = 0$  defines a set of states with no magnetic moment and a finite quadrupole moment. We refer to this set as  the set of quadrupolar states which include the state with zero eigenvalue for the $S_z$ operator. In this sense,  the observation of $\pi$ Berry phase in Ref.\cite{Robbins_1994} for  $S=1$ is an early example of quadrupolar Berry phase. 

We show that there exists a local gauge choice in which the quadrupolar states form a vector space defined over the field of real numbers. As geometric phase is independent of the choice of the local gauge, cyclic evolution in the quadrupolar subspace can lead to geometric phase which is quantized to zero or $\pi$ (as the only nontrivial phase comes from multiplying a state by $-1$ owing to reality constraints). In this article, we employ Majorana stellar representation (MSR) to arrive at a geometric visualization of the topological properties of the quadrupolar subspace. MSR is an old idea due to Ettore Majorana~\cite{Majorana1932}  which has gained prominence recently as a tool to understand geometric properties of finite-dimensional Hilbert spaces~\cite{Martin2010, Bruno2012,Ganczarek2012,Liu2014,Liu2016,Yao2017,Dogra_2018,Kam2021,Sandeep2022,Xingyu2022} and their connection to entalgement properties of permutation symmetric states of spin half system. The original idea of Majorana has also been extended to permutation symmetric states of higher spin systems\cite{Karol-spin1}.

This article is organized as following.
In section~\ref{sectionTopologicalSubspace}, we discuss a possible way to construct subspace of a Hilbert space which supports only zero or $\pi$ Berry phase and explore its connection to anti-unitary symmetries. 
In section~\ref{sectionQuadrupolarSubspace}, we introduce the quadrupolar subspace of a qutrit ($S=1$) and identify a global unitary map between the quadrupolar and real subspaces of the qutrit hence justifying the topological nature of the former. We also arrive at a connection between the quadrupolar subspace and the time-reversal symmetry. In section~\ref{sectionMSR}, we show that the exchange of Majorana stars leads to quantized Berry phase of $\pi$ for quadrupolar subspace.  We  discuss implications drawn from Majorana stars representations of eigenstates of quadrupolar Hamiltonians. We also discuss how the picture of exchange of Majorana stars evolves  as we interpolate between the quadrupolar subspace and the real subspace via a family of global unitary transformations.
In section~\ref{sectionDynamics}, we explore the influence of static Hamiltonian which keeps the states within quadrupolar subspace under time- evolution and discuss the geometric phase (of the Aharonov-Anandan type) accumulated by the quadrupolar state under cyclic time evolution.
In section~\ref{sectionSummary}, we summarize our findings.
\section{Topological subspace}\label{sectionTopologicalSubspace}
A subspace of a Hilbert space is considered to be topological if it allows only two values for the geometric phase ($0$ and $\pi$) for any closed loop in the corresponding ray space.
It may not be easy to determine whether a given subspace is topological without exhaustive explicit calculations.
However there is one well-known topological subspace for any quantum system - the set of all real states (in some chosen basis)~\cite{Samuel2001}. It is obvious that the phase accumulated by a state in the real subspace of states, due to closed loop evolution in the corresponding ray space can only be $0$ or $\pi$ (the state getting multiplied by $-1$) as all the states along the loop are real. Hence the real subspace qualifies as a topological subspace.

To explicitly verify this intuitive observation, let us consider the $N^{th}$ order Bargmann invariant\cite{Mukunda} defined as the cyclic inner product of states from this subspace given by 
$ \langle \psi_0    |\psi_1 \rangle
  \langle \psi_1    |\psi_2 \rangle \cdots
  \langle \psi_{N-1}|\psi_0 \rangle $,
where $|\psi_i\rangle$  for $i=0,1,..,N-1$ are $N$ distinct states with no two successive states being orthogonal in the sequence of states. In general this is a complex number whose argument is the geometric phase accumulated by the state corresponding to this closed loop projections. But for this subspace, each of the terms of the product are real numbers hence   resulting in the geometric phase being either $0$ or $\pi$ when the product is positive or negative respectively.
It is straightforward to construct examples with $0$ phase - we choose each state to have either all positive or all negative projections on a orthonormal basis.
Because each state appears twice in the product (as ket and bra) the cyclic product will be positive.
We now present one method to construct a negative cyclic product.
For any two states let us consider the geodesic loop passing through them~\cite{Mukunda1993a}.
If the two states are chosen to be from the real subspace then all the states in the geodesic loop will also be real.
In fact the geodesic is the one parameter subspace (over real field) spanned by the two chosen states.
It is straightforward to show that the cyclic inner product of $N$ equidistant (with respect to the parametrization angle) states on a geodesic parametrized by an angle in $[0,\pi)$ is
\begin{eqnarray}
        \langle \psi_0    |\psi_1 \rangle
  \langle \psi_1    |\psi_2 \rangle \cdots
  \langle \psi_{N-1}|\psi_0 \rangle 
=\left(\cos \frac{\pi}{N}\right)^{N-1}\cos\left(\pi-\frac{\pi}{N}\right), \nonumber\\
\end{eqnarray}
which is real and negative. 
The above construction is a general one and true for any loop  comprising a shorter and longer geodesic connecting two distinct nonorthogonal state.
Its relevance here is just because the geodesic loop passing through two chosen real states lies within the real subspace.
This property has been used here to provide an explicit construction of loops within the real subspace with $\pi$ geometric phase.
From the sets of all real states and corresponding real 
Hamiltonians ( which maps the set of real states back to itself ), one can generate a continuous family of sets with exactly same geometric and topological properties by applying family of global unitary transformations on them. We will show that the quadrupolar subspace discussed here is one such example of a subspace which is connected to the real subspace by a global unitary transformation.

\subsection{Invariance of states of a topological subspace under an anti-unitary operator} \label{InvariantSubspace}
There is another way to characterize such topological subspaces without explicitly referring to the mapping to real subspace.
States from the real subspace are invariant under the anti-unitary operation of complex conjugation.
In fact this invariance can be taken to be the defining property of the real subspace\cite{Hatsugai}. We now study such invariances in other subspaces which are unitarily connected to the real subspace.
Let us denote the complex conjugation by $\mathcal{K}$ and consider a arbitrary state $|{\psi}\{x_1,x_2,\cdots x_n\} \rangle_R$ from the largest real subspace of a finite dimensional Hilbert of dimension $n$ where subscript $R$ stands for the real subspace and the set $\{x_1,x_2,\cdots x_n\}$ represents the $n$ real parameters required to parametrize the state such that,
\begin{equation}
\mathcal{K} |\psi \rangle_R = |\psi \rangle_R.
\end{equation}
Let us now apply an arbitrary unitary transformation $U$ on both sides
\begin{eqnarray}
    \label{anti-unitary-1}
U |\psi \rangle_R = U \mathcal{K} |\psi \rangle_R
= \mathcal{K} U^* |\psi \rangle_R
&=& \mathcal{K} (U^* U^\dagger) U |\psi \rangle_R \nonumber\\
&=& \Theta U |\psi \rangle_R,
\end{eqnarray} 
where $\Theta = \mathcal{K} (U^* U^\dagger)$ is another anti-unitary operator and $U |\psi \rangle_R$ is invariant under it. Thus we see that all topological subspaces unitarily related to the real one are made up of states that are invariant under an anti-unitary operator. It is important to note that the state  $U |\psi\{x_1,x_2,\cdots x_n\}\rangle_R$ is in general defined up to a $U(1)$ phase which can be different for different values of $x_1,x_2,\cdots x_n$ and hence it will be difficult to check this invariance when the states from the subspace are expressed in a arbitrary $U(1)$ gauge. One way to get around it is to check if it is possible to make a new $U(1)$ gauge choice for the subspace under consideration, such that the subspace in this new gauge choice forms a vector space over the field of real numbers.

Now let us note some properties of the anti-unitary operator $\Theta$ -
\begin{equation} \label{anti-unitary-2}
\Theta^2 = \mathcal{K} (U^* U^\dagger) \mathcal{K} (U^* U^\dagger)
= \mathcal{K} U^* U^\dagger U U^T \mathcal{K} = 1.
\end{equation} 
We also note that our construction of anti-unitary operator involves symmetric unitary operators
$$(U^* U^\dagger)^T = (U^\dagger)^T (U^*)^T = U^* U^\dagger.$$
Since any symmetric unitary operator can be written as $U^* U^\dagger$ we have in fact obtained a general prescription to define the topological subspaces unitarily related to the real one.
Such a subspace consists of states that are invariant under the action of an anti-unitary operator made up of a symmetric unitary operator followed by complex conjugation.
We note here that the invariance under an anti-unitary operator discussed above is completely equivalent to the real mapping and in fact the same unitary operator is used in the construction of the anti-unitary operator characterizing the subspace.
As mentioned before, there is a continuous family of topological subspaces which are related to each other and the real subspace by global unitary transformations.
The quadrupolar subspace is just one of them for the three-level system.
We next discuss the quadrupolar subspace and its characterization  using Majorana stellar representation.
\section{Quadrupolar subspace} \label{sectionQuadrupolarSubspace}
We now specialize the general ideas of the previous section to the physically relevant case of the quadrupolar subspace of a three-level or spin-1 system (qutrit).
This subspace is defined as the set of states with zero magnetization in all directions
\begin{equation} \label{QuadrupolarStates}
\langle \psi | \vec{S} | \psi \rangle = 0.
\end{equation}

The topological nature of the quadrupolar subspace can be established by realizing that the entire subspace can be made real by a global unitary transformation.
To show this let us start from the most general three-level state
\begin{equation} \label{GeneralState}
|\psi \rangle = \left[\begin{matrix}\alpha\\\beta\\\gamma\end{matrix}\right]
\end{equation}
Upon imposing Eq.\ref{QuadrupolarStates} and choosing $\beta$ to be real and positive (which can be thought of a freedom of choosing an overall phase for the state, which do not change the physical state), we get the normalized quadrupolar states as
$$|\psi_{q} \rangle = \left[\begin{matrix}\alpha\\\beta\\-\alpha^{*}\end{matrix}\right]\quad \text{where},\ \  2|\alpha|^{2}+\beta^{2}=1$$
With this parametrization it is staightforward to see that the quadrupolar states form a vector space over the field of real numbers. We may also choose a Bloch sphere like parametrization as follows
$$|\psi_{q} \rangle = \left[\begin{matrix}\frac{1}{\sqrt{2}} e^{i\phi}\sin(\theta/2)\\ \cos(\theta/2) \\ \frac{-1}{\sqrt{2}}e^{-i\phi}\sin(\theta/2)\end{matrix}\right]$$
where $\theta\in\left[0,\pi\right]$ and $\phi\in\left[0,2\pi\right)$. With this parametrization each point on the surface of a sphere corresponds to a quadrupolar state.
Now consider the effect of the following unitary transformation
\begin{equation} \label{GlobalUnitary}
U = \frac{1}{\sqrt{2}} \left[\begin{matrix}i & 0 & i\\0 & \sqrt{2} & 0\\1 & 0 & -1\end{matrix}\right]
\end{equation}
on the quadrupolar states,
$$U |\psi_q \rangle = \frac{1}{\sqrt{2}} \left[\begin{matrix}i(\alpha-\alpha^{*})\\ \sqrt{2}\beta\\(\alpha+\alpha^{*}) \end{matrix}\right],$$
which makes the quadrupolar states real. Same can be verified with the Bloch sphere like parametrization.
Since all the states have become real by a single unitary transformation, the quadrupolar subspace has the same geometric and topological properties as the real subspace which is topological.
Let us also construct the anti-unitary operator that leaves states from quadrupolar subspace invariant
$$\Theta = \mathcal{K} U_q^* U_q^\dagger = \mathcal{K} \left[\begin{matrix}0 & 0 & -1\\0 & 1 & 0\\-1 & 0 & 0\end{matrix}\right]$$
where $U_q$ is the unitary operator that maps real subspace to the quadrupolar one, i.e. inverse of Eq.\ref{GlobalUnitary}. We can explicitly check the invariance of the quadrupolar states under this anti-unitary operator as 
$$\Theta\left|\psi_q\right\rangle=\mathcal{K}\left[\begin{matrix}0 & 0 & -1\\0 & 1 & 0\\-1 & 0 & 0\end{matrix}\right]\left[\begin{matrix}\alpha\\\beta\\-\alpha^{*}\end{matrix}\right]=\mathcal{K}\left[\begin{matrix}\alpha^{*}\\\beta\\-\alpha\end{matrix}\right]=\left|\psi_q\right\rangle
$$
\subsection{Time reversal symmetry}\label{subsectionTRS}
In quantum theory, complex conjugation is intimately related to the time-reversal operation and indeed there is another way to obtain the above anti-unitary operator for the quadrupolar subspace.
A spin-1 system can be considered as the triplet sector of two spin-1/2 systems.
The time-reversal operator for two spin-1/2 systems, $\mathcal{T}_{2\times 2} = \mathcal{K} (i\sigma_y \otimes i\sigma_y)$, upon projecting to the triplet sector gives
$$\mathcal{T}_{2\times 2}^\mathrm{tr} = \mathcal{P} \mathcal{T}_{2\times 2} \mathcal{P}^\dagger = \mathcal{K} \left[\begin{matrix}0 & 0 & 1\\0 & -1 & 0\\1 & 0 & 0\end{matrix}\right]$$
where $\mathcal{P}$ is the projection operator from the $2\otimes 2$ Hilbert space to the triplet sector.
Thus the anti-unitary operator of our construction is the exact negative of the time-reversal operator in the triplet sector.
Hence the anti-unitary operator that leaves states from the quadrupolar subspace invariant is just time-reversal.
We now explicitly show that this is in fact the defining property of quadrupolar subspace.
Acting the time-reversal operator on a general state Eq.~\ref{GeneralState} from the triplet sector we get
$$\mathcal{T}_{2\times 2}^\mathrm{tr} |\psi \rangle = \mathcal{K}\left[\begin{matrix}\gamma\\ -\beta\\\alpha\end{matrix}\right]= \left[\begin{matrix}-\gamma^{*}\\ \beta^{*}\\-\alpha^{*}\end{matrix}\right].$$
Demanding $\mathcal{T}_{2\times 2}^\mathrm{tr}|\psi \rangle=-|\psi \rangle$ results in the exact conditions defining the quadrupolar subspace i.e. $\alpha=-\gamma^{*}$ and $\beta=\beta^{*}$. 
With this we also note that in general, the real space does not have time-reversal symmetry as by defining property they are invariant under complex-conjugation.

\section{Majorana stellar representation} \label{sectionMSR}
We present a geometric way to study the topological nature of the quadrupolar subspace via the MSR.
In MSR the state of a $d$-dimensional system is represented by $d-1$ points or stars on a Bloch sphere.
These stars are identical in the sense that interchanging them does not change the state they correspond to.
 The indistinguishability of stars gives another physical meaning to this representation - these $d$-level states are completely symmetrized states of $(d-1)$ spin-1/2 systems\cite{bloch_rabi_1945}.
These symmetrised states are nothing but a spin-$(d-1)/2$ state.  
Using this symmetrization, $d$-dimensional Hilbert space is mapped to a polynomial vector space of degree $(d-1)$ over complex field.
A generic state of spin-$S=(d-1)/2$ is $|\psi_{S}\rangle=\sum_{m=-S}^{S}c_{m}|S,m\rangle$ mapped to the Majorana polynomial(MP) $\sum_{r=0}^{2S}a_{r}x^{2S-r}$ with\cite{Majorana1932}
\begin{equation}\label{MajoranaFormula}
    a_{r}=(-1)^{r}\frac{c_{s-r}}{\sqrt{(2S-r)!r!}}
\end{equation}
The MP has $(d-1)$ roots, which can be written as $\tan\frac{\theta_{k}}{2} e^{i\phi_{k}}$ with $k\in\{1,2,...,d-1\}$.
Then the spherical coordinates $(\theta_{k},\phi_{k})$ gives the unit vector $\hat{u}_{k}$, the location of MSs on the Bloch sphere. 
The MSR of quadrupolar states turns out to be particularly simple---a pair of stars that are antipodal(see appendix \ref{appendix_antipodal_MSs}).
It is easy to understand the MSR for the quadrupolar states---as the quadrupolar states have zero magnetization they must correspond to two spin-1/2 particles with opposite magnetization and hence must have antipodal stars.
Many properties of the quadrupolar subspace can be immediately demonstrated in the MSR, such as time-reversal symmetry of the states.
Because these states have zero magnetization they possess time-reversal symmetry which is manifest in the MSR, since the time-reversal operation in MSR amounts to taking every star to its antipodal location.
Another non-trivial property of this subspace is that the entanglement between two constituent spin-1/2 particles is maximum.
Again this is simple to see in the MSR as all states are formed by the symmetrized linear combination of states which are antipodal (and hence orthogonal) and as a result are of the form $(|01 \rangle + |10 \rangle) / \sqrt{2} $ in suitable basis, one of the Bell states having maximum entanglement.
\begin{figure}
\includegraphics[clip,trim=0mm 70mm 0mm 0mm,width=\columnwidth]{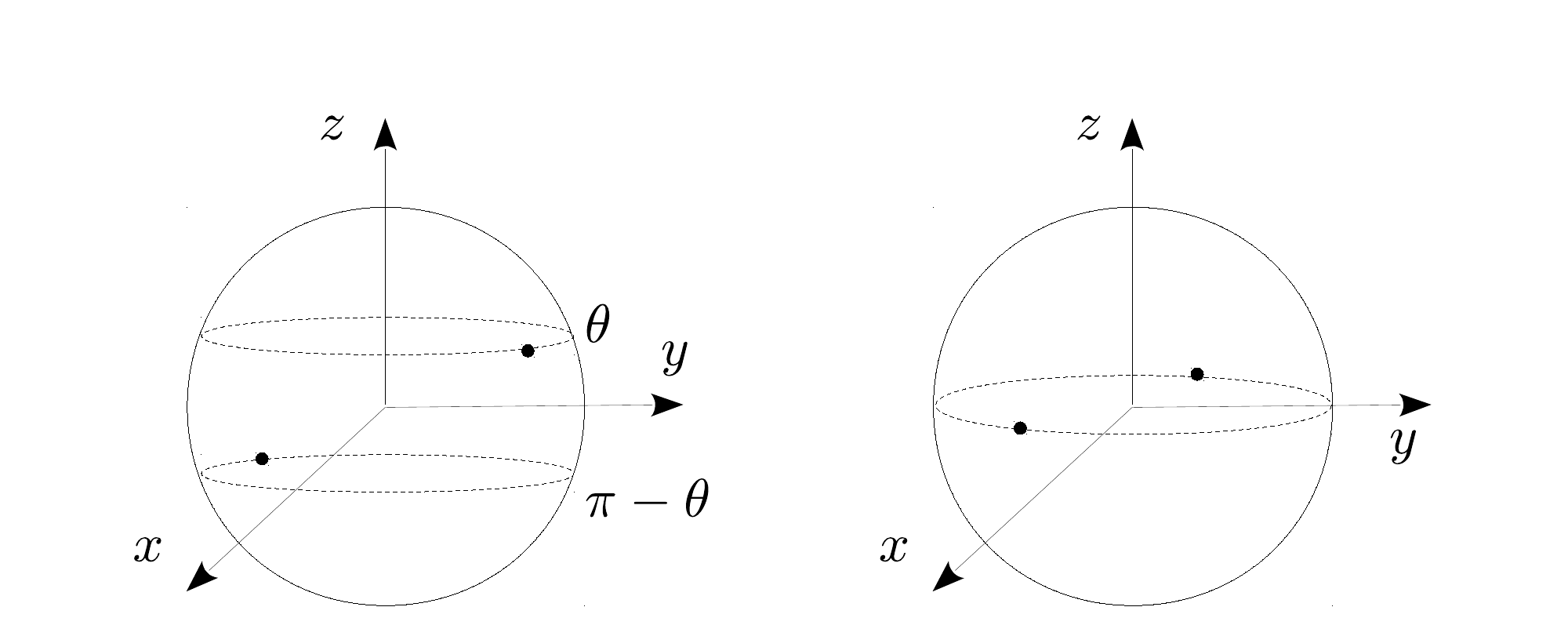}
\caption{MSR of states from the quadrupolar subspace.}
\label{QuadrupolarLoops}
\end{figure}

\subsection{Berry phase in MSR}\label{4A}
Understanding geometric phase in terms of MS is a well studied topic~\cite{Bruno2012,Liu2014}. When a cyclic change of states result in individual closed loop trajectories for each MSs representing the state, the corresponding Berry phase for the state is   given by\cite{Liu2014}
\begin{equation}\label{BerryPhaseInMSR}
    \gamma^{(d)}=\gamma^{(d)}_{0} + \gamma^{(d)}_{C}~,
\end{equation}
which is a sum of two contributions - one part is $\gamma_{0}^{(d)}=-\sum_{i=1}^{d-1} \Omega_{i} / 2$, the sum of solid angle subtended by each star (modified by a $\pm$ sign depending on whether the loop points towards or away from the origin in the right hand screw rule sense) and the other part is 

\begin{equation}\label{eq:2}
\gamma_{C}^{(d)}=\frac{1}{2} \oint \sum_{i=1}^{d-1} \sum_{j(>i)}^{d-1} \beta_{i j} \Omega\left(\mathrm{d} \hat{u}_{i j}\right)
\end{equation}
Here, $\beta_{i j}$ is the correlation factor given by
\begin{equation}
\beta_{i j}(\boldsymbol{D}) \equiv-\frac{d_{i j}}{N_{d-1}^{2}(\boldsymbol{D})} \frac{\partial N_{d-1}^{2}(\boldsymbol{D})}{\partial d_{ij}}
\label{correlation}
\end{equation}
with $\boldsymbol{D}=\{d_{ij}\},\ i<j$; $d_{ij}=1-\hat{u}_{i}.\hat{u}_{j}$ and $N_{d-1}^{2}(\boldsymbol{D})$ is the normalization coefficient of state $|\psi_S\rangle$ given in terms of $\hat{u}_{i}$'s.
The term $\Omega\left(\mathrm{d} \hat{u}_{i j}\right) \equiv \hat{u}_{i} \times \hat{u}_{j} \cdot d\left(\hat{u}_{j}-\hat{u}_{i}\right) / d_{i j}$ is the sum of solid
angles of the infinitesimally thin triangle $\left(\hat{u}_{i},-\hat{u}_{j},-\hat{u}_{j}-d \hat{u}_{j}\right)$
and $\left(\hat{u}_{j},-\hat{u}_{i},-\hat{u}_{i}-d \hat{u}_{i}\right)$.
It can be interpreted as the solid angle due to the relative motion between each pair of stars and their absolute evolution~\cite{Liu2014}.
This prescription works when the cyclic evolution of the state gives rise to cyclic evolution of each of the individual stars.
As mentioned earlier, since the stars are identical it is possible that the cyclic evolution of the state may result in a permutation of a fewer stars\cite{Kam2021} in which case these stars will not complete a closed loop individually.
Hence, in the case when the cyclic evolution of the states result in permutation alone (i.e. no individual cyclic loops for any star), it corresponds to a possibility which is beyond the discussion presented in Ref.~\onlinecite{Liu2014}.
If the permutation is such that there is effectively just one loop, the Berry phase is the solid angle of this loop.
We will see below that the quadrupoalr case of interest to us has this property.
\subsection{MSR of quadrupolar subspace}
The fact that quadrupolar states are represented by antipodal stars implies that we need to keep track of only one of the stars as it immediately tells us the location of the other. This is a very convenient situation geometrically as the complexity of visualizing this subspace is the same as that of a spin-1/2 system, namely everything happens on the surface of a Bloch sphere and a state is represented by a point. But  topology of the two problems are quite distinct as discussed in details in Ref.~\onlinecite{Robbins_1994}. Actually this geometric representation using the MSR clearly show that the space of quadrupolar states lay on the Bloch sphere with antipodal point being identified hence providing yet another route to arrive at the conclusion that the topology of parameter space for the quadrupolar subspace is the real projective plane ($RP^2$)\cite{Pollmann2014}. 

Now we will show that the MSR provides an elegant way to visualize the quantization of Berry phase to zero or $\pi$ for quadrupolar states. We start by noting that the closed loops in ray space of quadrupolar subspace can be of two types - ones where the two stars individually make loops and others where they get exchanged.
Because the two stars are always antipodal there is no relative motion between them and hence there is no correlation contribution ($\gamma^{(d)}_{C}=0$) to the Berry phase, i.e., $\gamma^{(d)}=\gamma^{(d)}_{0}$. In fact the correlation contribution for the real subspace is also zero(see appendix \ref{correlation_term}) but unlike the quadrupolar case it is not easy to visualize geometrically.  

In the first case when the two stars make individual loops and  their contributions exactly cancel each other [a simple example of such situation is shown in Fig.\ref{QuadrupolarLoops}(a)] and hence $\Omega_{1}=-\Omega_{2}$, resulting in zero Berry phase.
In cases when the stars exchange their location [see e.g. Fig.\ref{QuadrupolarLoops}(b)] the combined trajectory subtends exactly $2\pi$ solid angle at the center giving rise to a Berry phase of $\pi$, i.e., $\gamma_{0}^{(d)}=\pi$. Not that the contributions due to individual trajectory of the Majorana stars to $\gamma_{0}^{(d)}$ by themselves do not corresponds to any gauge invariant geometric phase and hence $\gamma_{0}^{(d)}$ can not be expressed as a sum of solid angles(see appendix \ref{MS_Exchange}).
\begin{figure}
    \centering
    \includegraphics[width=5cm]{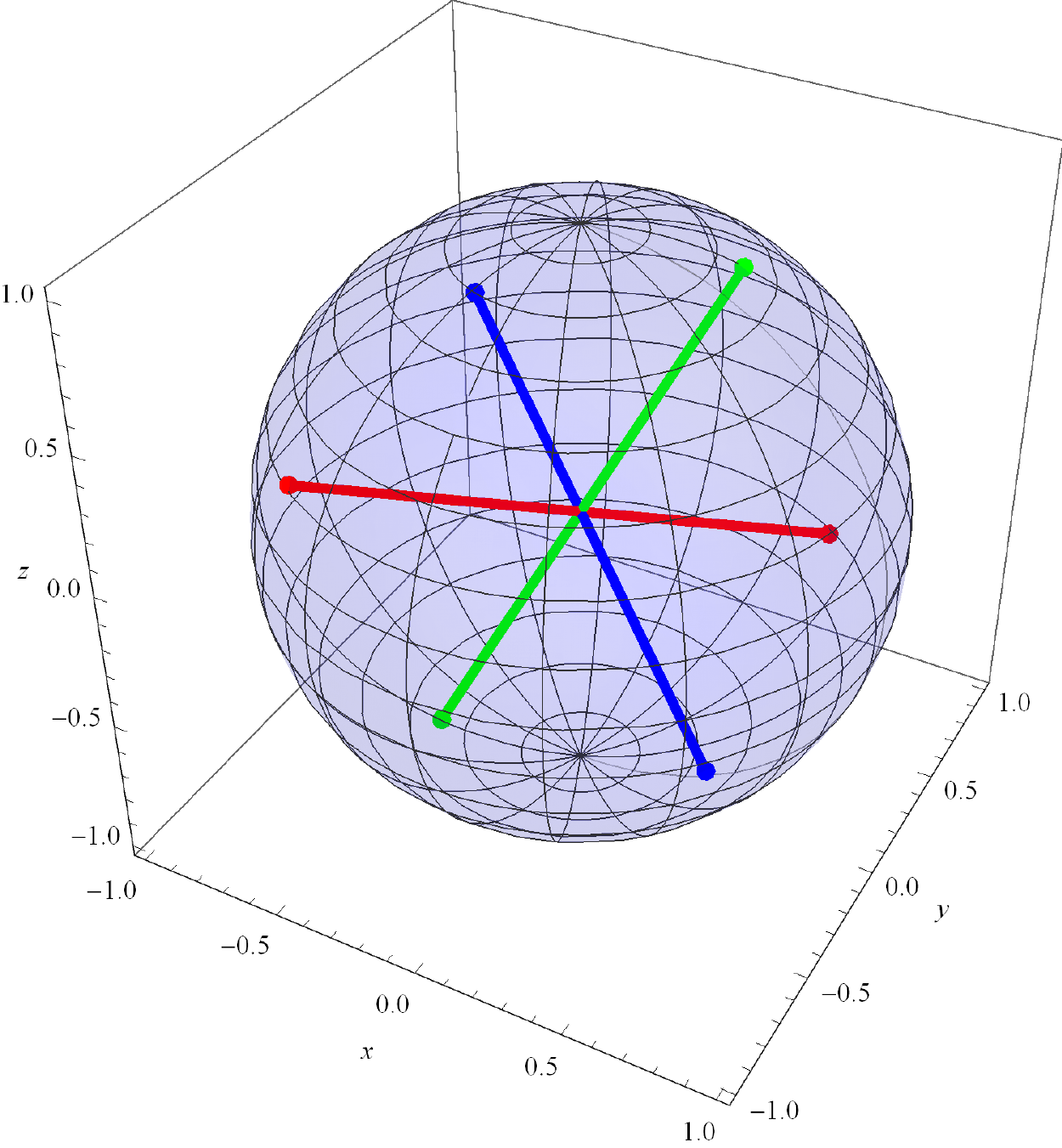}
    \caption{Geometry of MSs for quadrupolar eigenstates - the three orthogonal states are represented by three pairs of MSs along three mutually perpendicular lines in three different color.}
    \label{orthogonalsticks}
\end{figure}
It is not difficult to see that one of these two scenarios will hold for all possible loops and since we have exhausted all the possibilities for the quadrupolar subspace, we can conclude that this subspace is indeed topological.
With this understanding of quadrupolar states in terms of MSs, we can make interesting prediction about the topological properties of the eigenstates of quadrupolar Hamiltonian given by 

\begin{equation} \label{QuadrupolarHamiltonian}
H = \sum_{ij} \alpha_{ij} Q_{ij},
\end{equation}
where $Q_{ij}$ are components of the quadrupole moment tensor operator\cite{Maarten} expressed as 
\begin{equation} \label{QuadrupolarTerms}
Q_{ij} = \frac{1}{2} (S_i S_j + S_j S_i) - \frac{1}{3} S^2 \delta_{ij},
\end{equation}
 and  $S_i$ represents component of the spin operator along the  $i^{th}$ direction where $i=x,y,z$. As $Q$ represents a traceless symmetric tensor, hence ($Q_{ij} = Q_{ji}$) and the diagonal elements sum to zero ($Q_{xx} + Q_{yy} + Q_{zz} = 0$). Owing to this fact, only five components of quadrupole moment tensor operator are linearly independent and conventionally they are taken to be  $Q_{xy}, Q_{yz}, Q_{zx}$, $Q_{zz}$ and $Q_{xx} - Q_{yy}$. It is straightforward to check that the eigenstates of a quadrupolar Hamiltonian given in Eq.~\ref{QuadrupolarHamiltonian} belong to the quadrupolar subspace of states defined via Eq.~\ref{QuadrupolarStates}. Hence the  set of orthonormal quadrupolar eigenstates of such a Hamiltonian will occupy the six end points of a Cartesian coordinate system on the Bloch sphere in MSR as shown in Fig.\ref{orthogonalsticks}.
\begin{figure*}[t]
 \center
    \subfigure[]{\includegraphics[width=4cm]{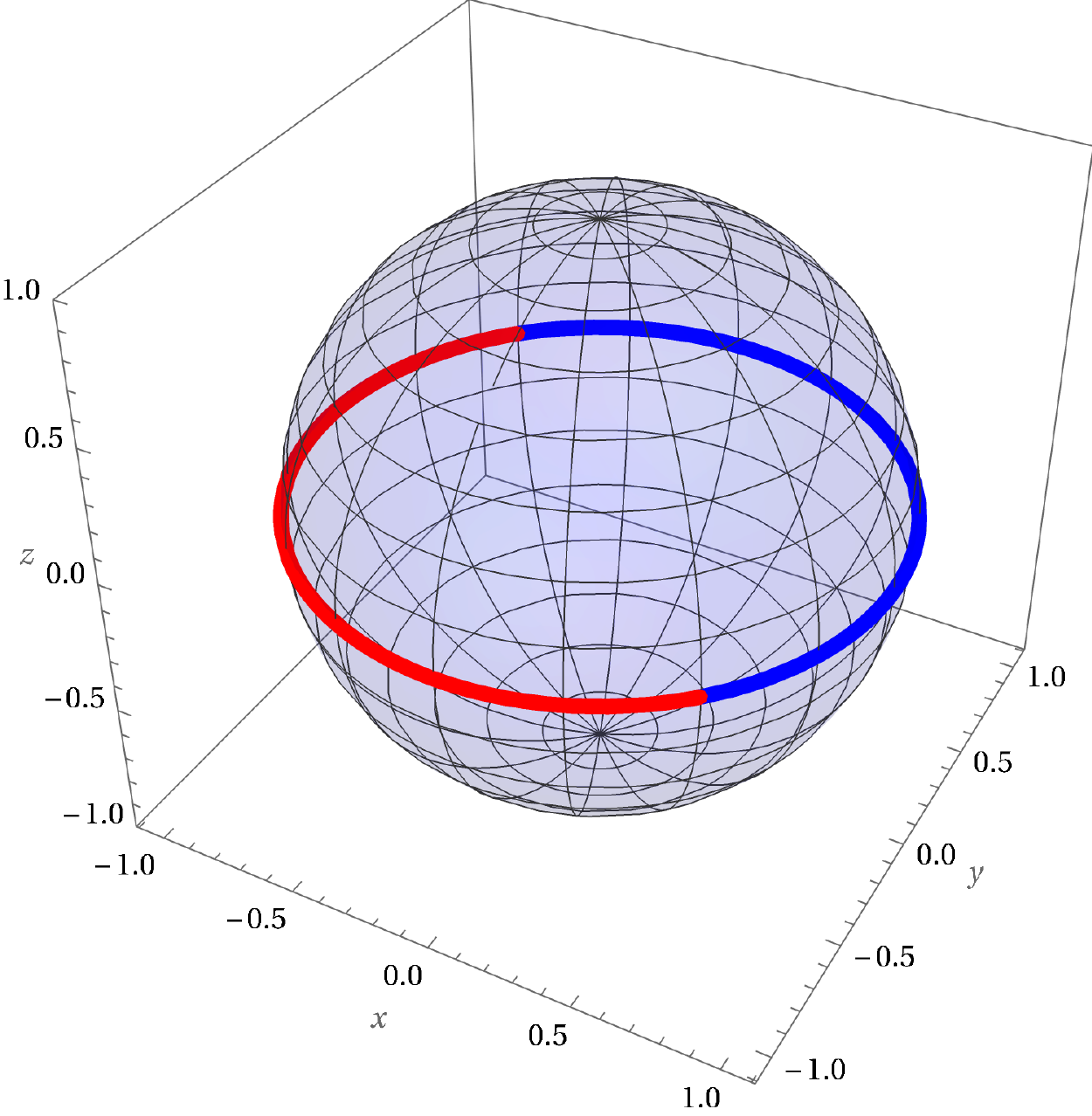}}
    \subfigure[]{\includegraphics[width=4cm]{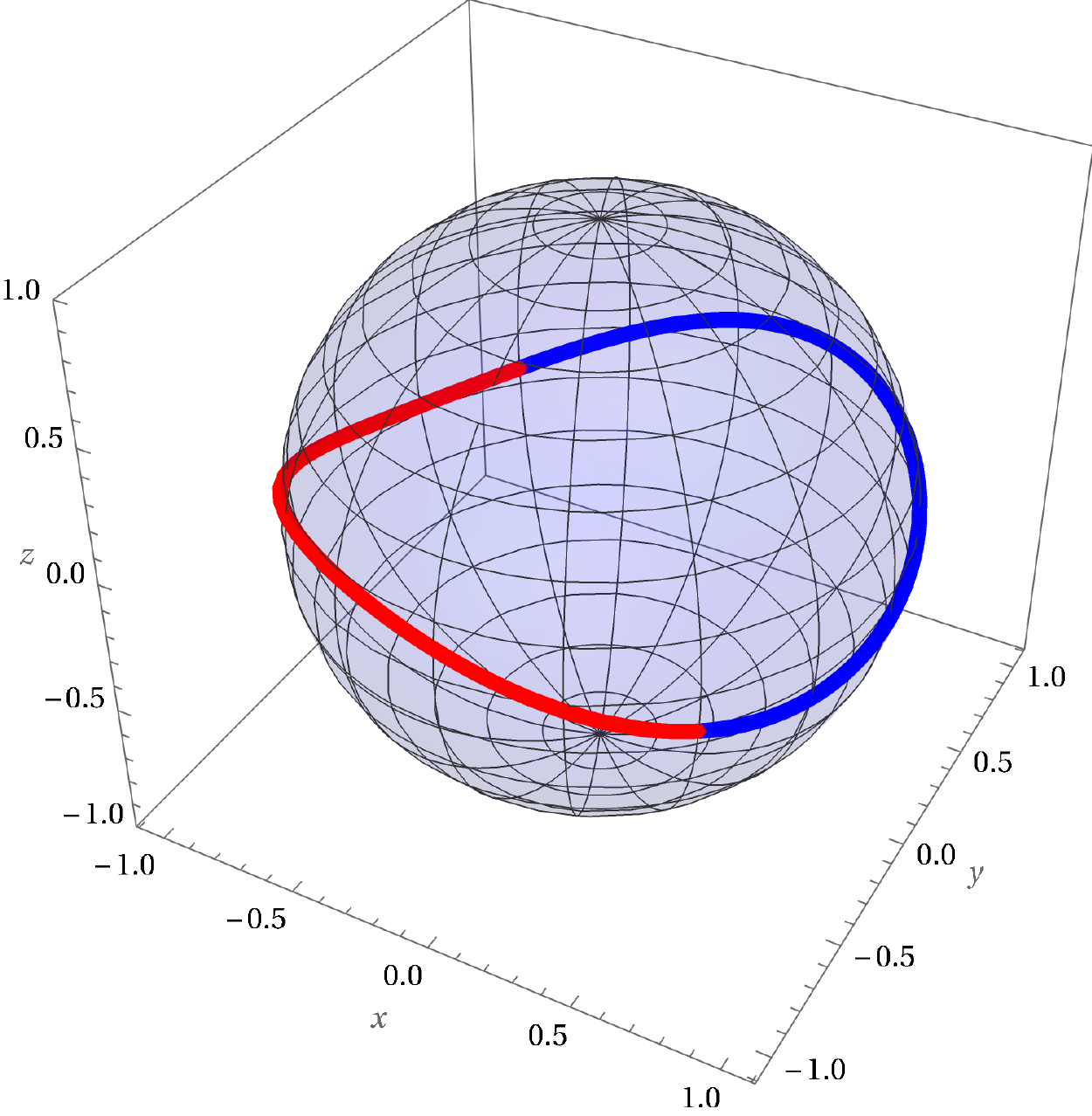}}
    \subfigure[]{\includegraphics[width=4cm]{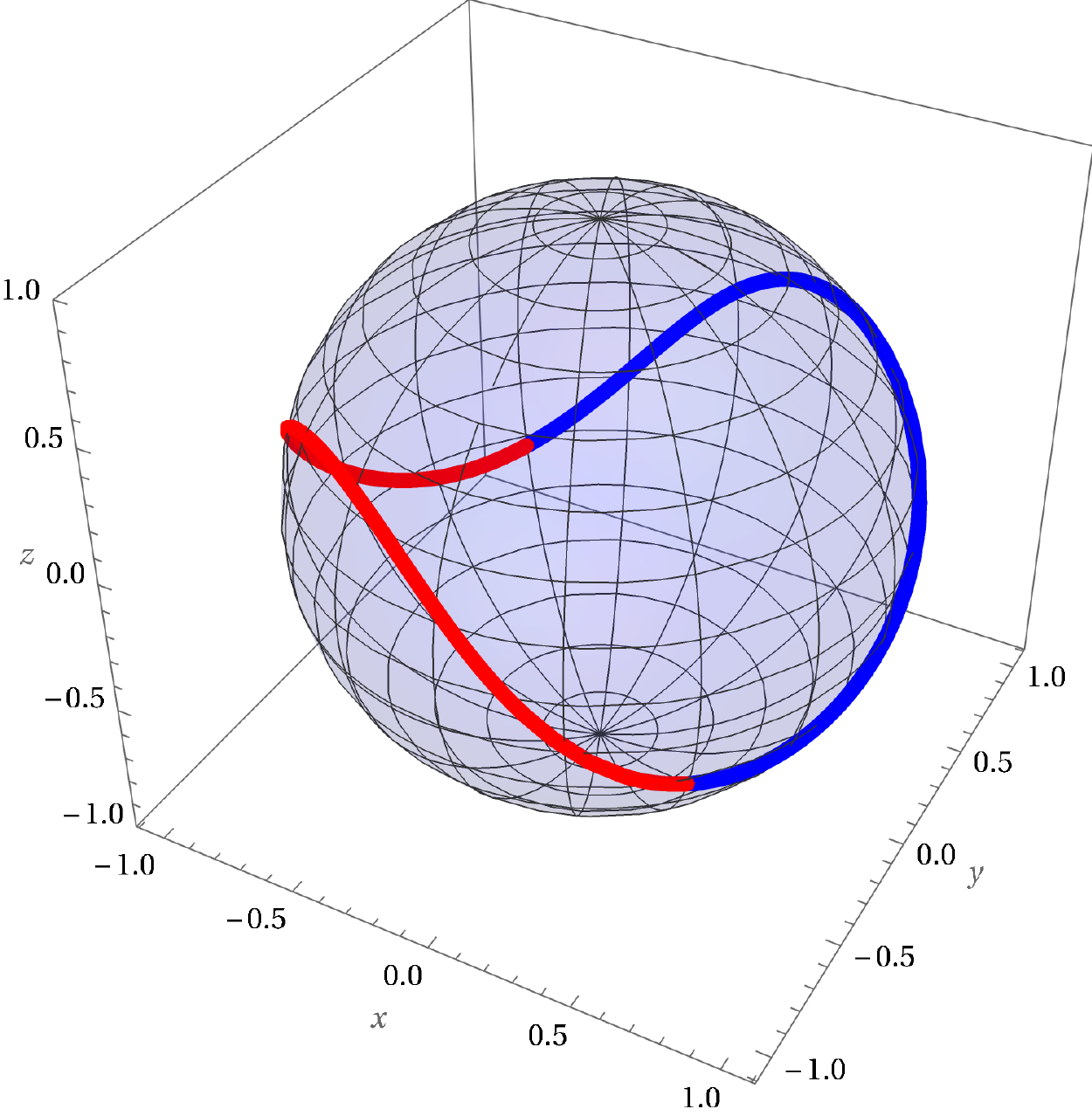}}
    \subfigure[]{\includegraphics[width=4cm]{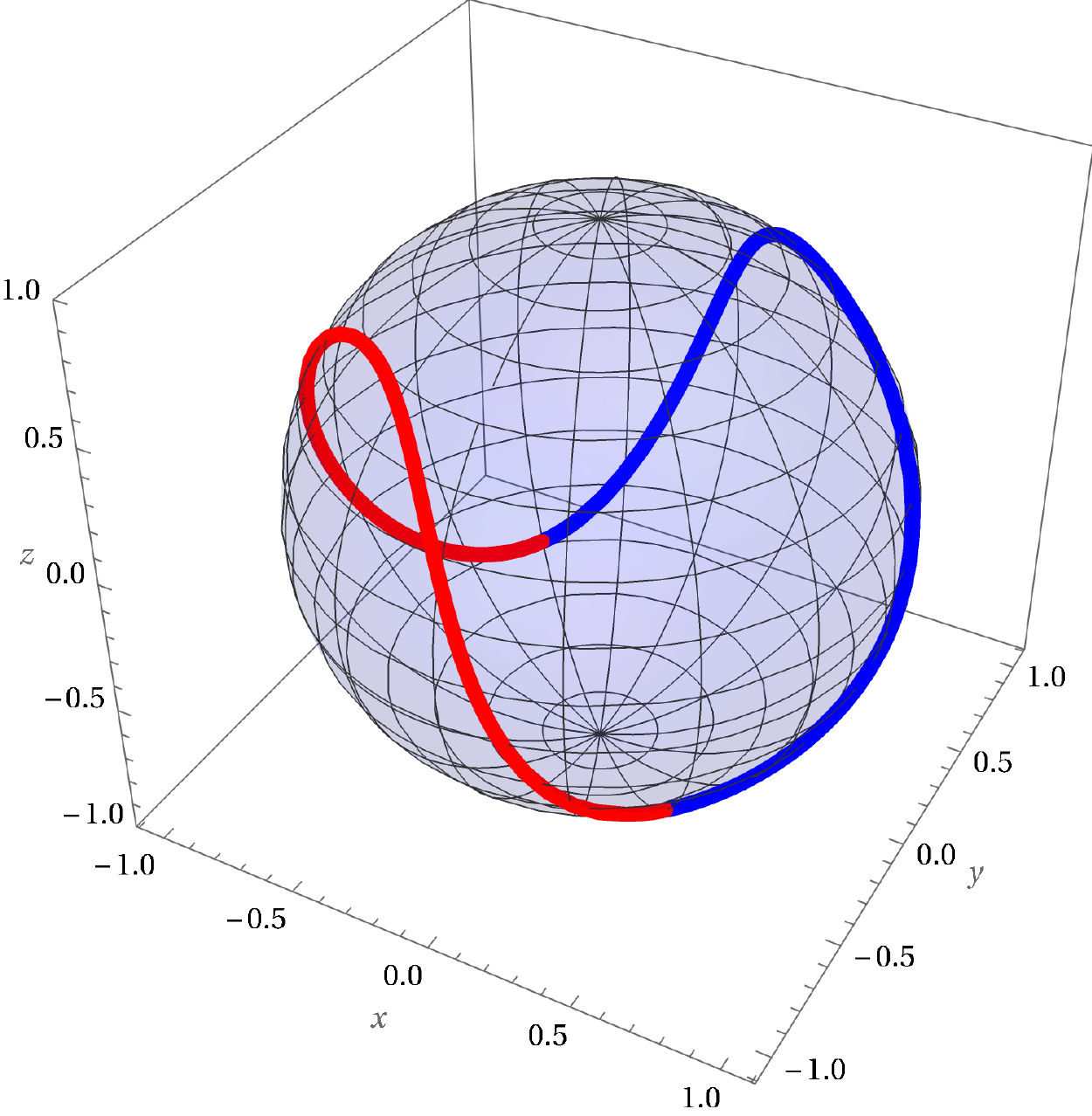}}
    \subfigure[]{\includegraphics[width=4cm]{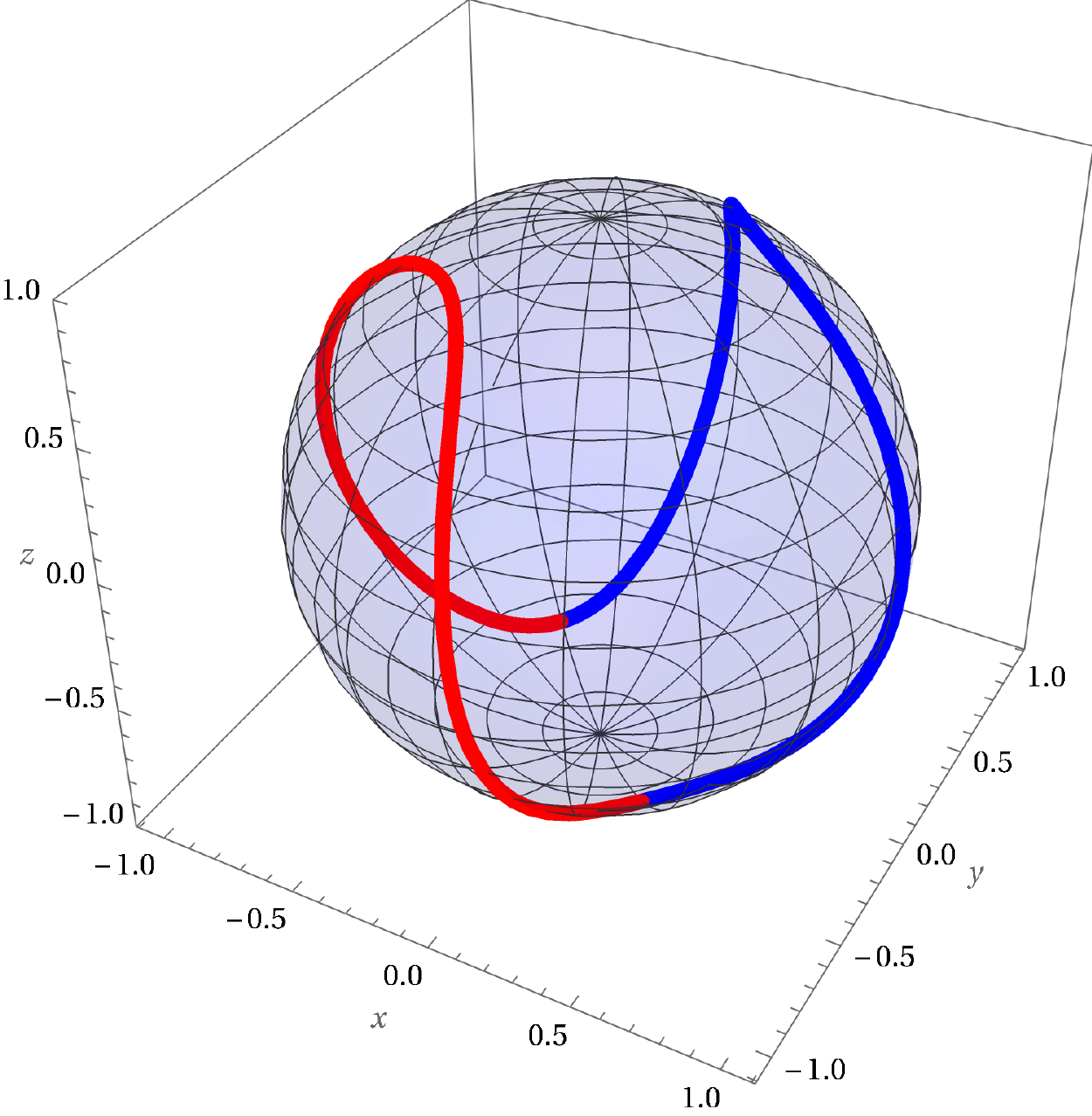}}
    \subfigure[]{\includegraphics[width=4cm]{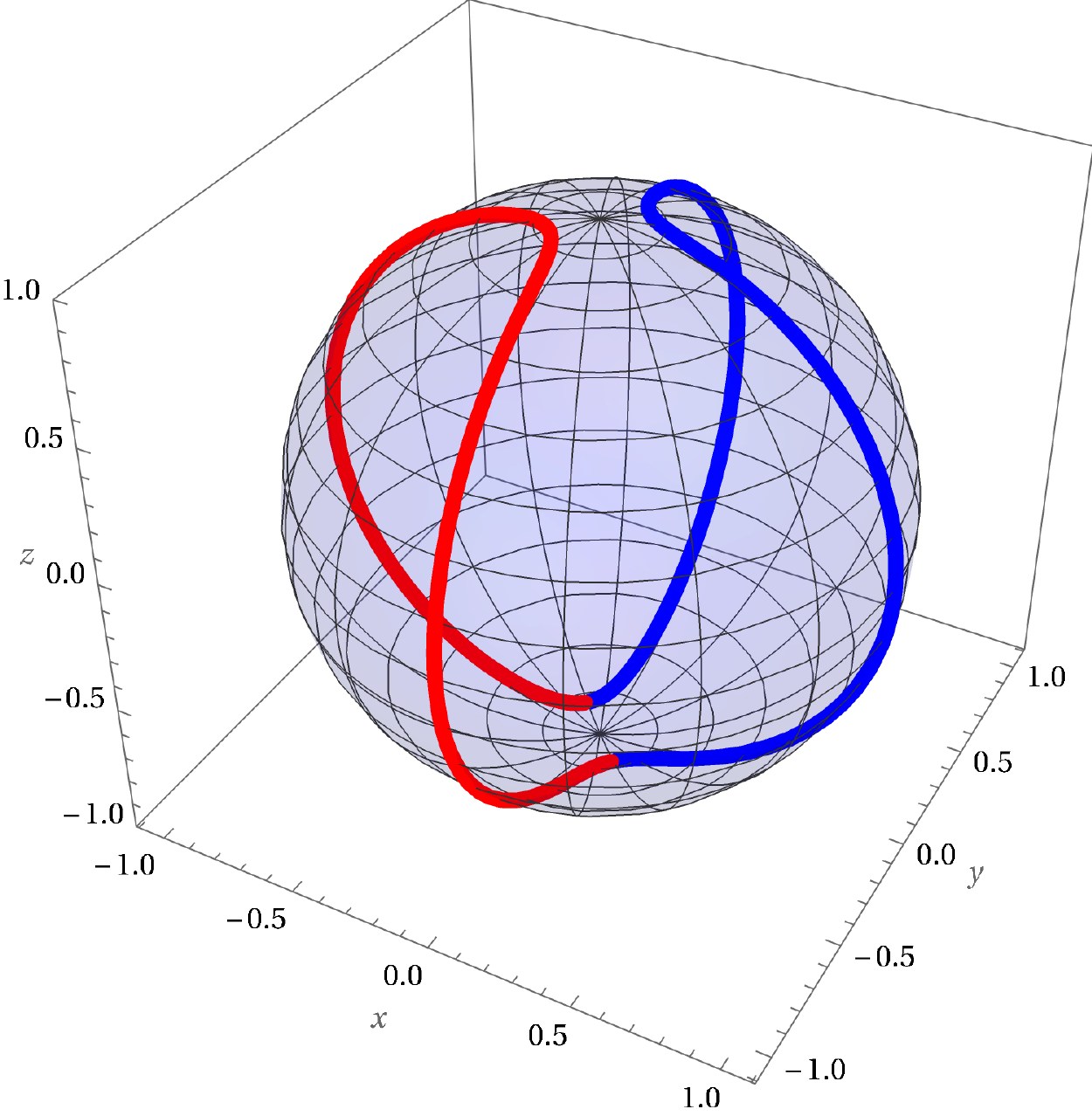}}
    \subfigure[]{\includegraphics[width=4cm]{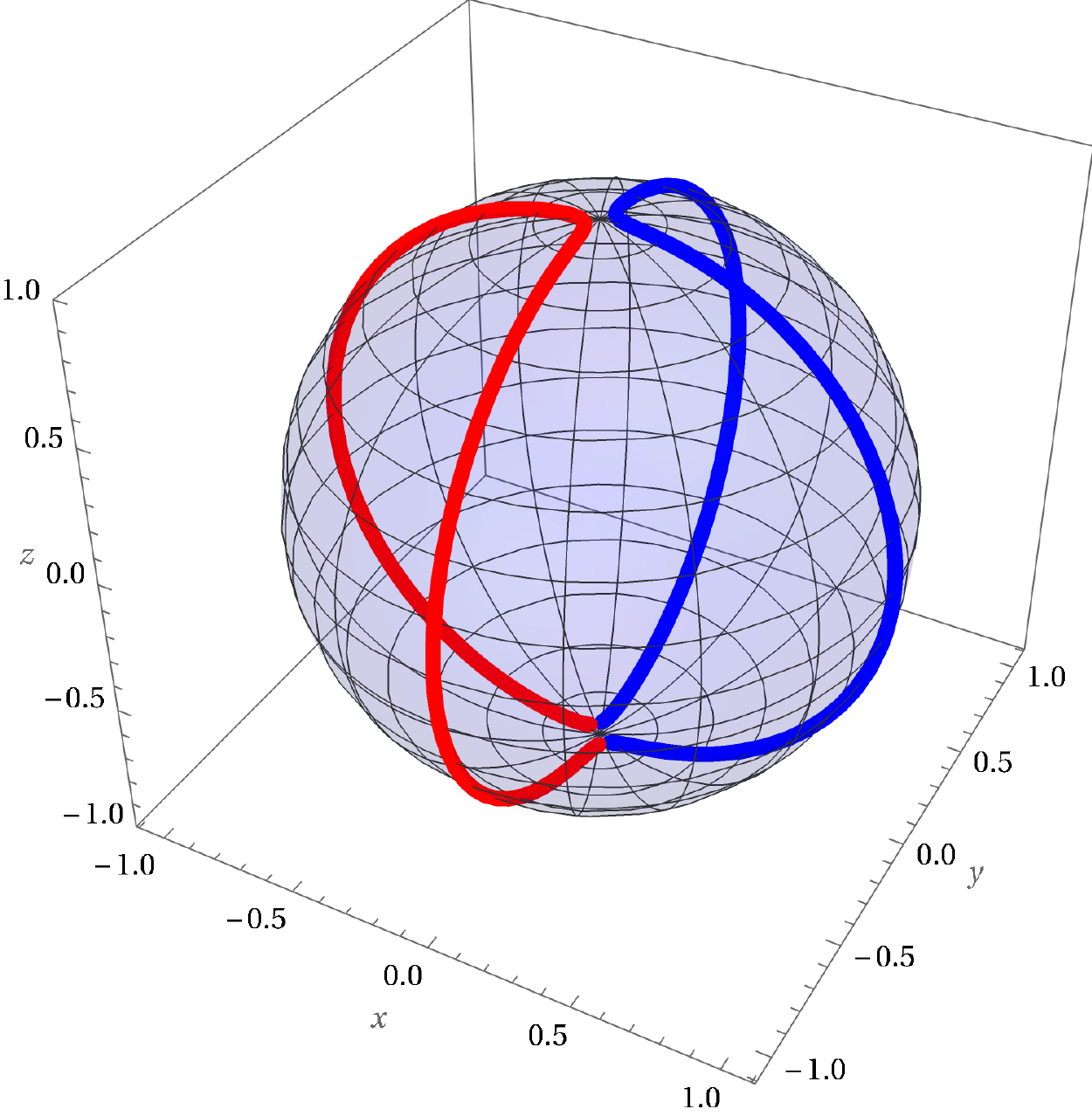}}
    \subfigure[]{\includegraphics[width=4cm]{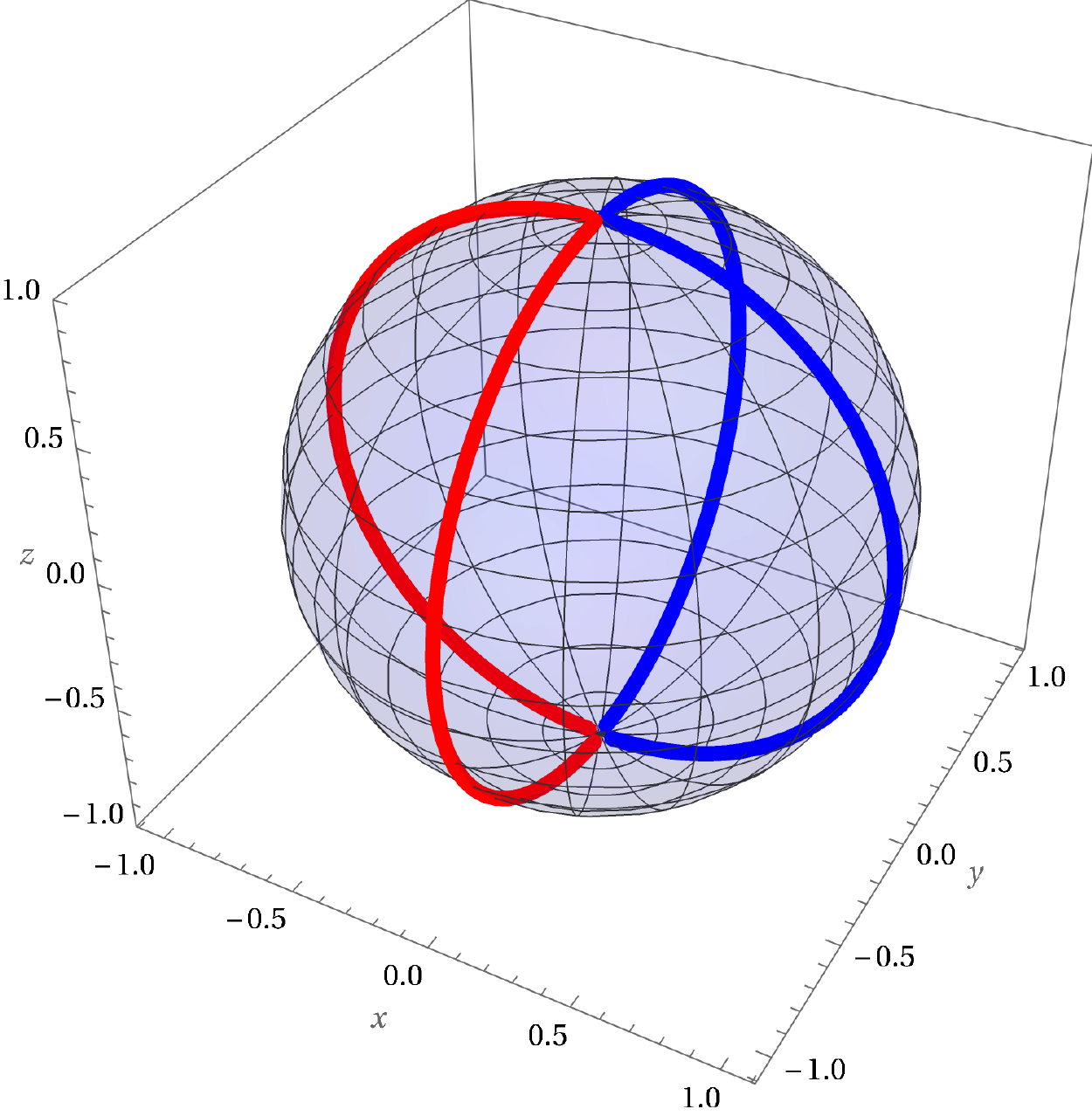}}

   \caption{Trajectories of MSs corresponding to an eigenstate of the Hamiltonian $H = U^{\dagger}(\alpha)(\cos\theta Q_{x^2-y^2} + \sin\theta Q_{xy})U(\alpha)$ (associated with the eigenvalue $\sqrt{5 +3\cos(2\theta})/2\sqrt{2}$)  as the parameter $\alpha$ of the unitary operator is changed from 0 to 1 (quadrupolar space to real space) shown in alphabetical order. The values of the parameter shown above are $\alpha$ = 0, 0.2, 0.5, 0.75, 0.9, 0.99, 0.999, 1 in same order as in the diagram. }
    \label{Parametrizedfig}
\end{figure*}
This result provides a simple geometric way to understand some topological properties of the system, such as it is not possible to have a single eigenstate which under adiabatic and cyclic time evolution of the Hamiltonian leads to a geometric phase of $\pi$. We will call these eigenstate as topological eigenstates.
In general a cyclic evolution of the quadrupolar Hamiltonian will correspond to a continuous rigid rotation of the six stars on the Bloch sphere. 
All eigenstate being topological requires all pair of antipodal MS getting exchanged which is equivalent to the inversion operation in three-dimension, which can not be obtained by any rotation. This implies that all the three eigenstates cannot be simultaneously topological. Also continuous rigid rotation can lead to exchange of a pair of MS and hence topological eigenstate can only appear in pairs.
Thus for a cyclic evolution of quadrupolar Hamiltonians we will either have no topological  eigenstates(trivial) or a pair of topological states(non-trivial).

\subsection{Mapping from quadrupolar to real subspace}
It is clear from the discussion above that Berry phase for the quadrupolar subspace and the real subspace, evaluated using the MSR given by Eq.~\ref{BerryPhaseInMSR} share the common feature that for both case the correlation terms vanishes, i.e., $\gamma^{(d)}_{C}=0$. Another commonality between the two is, for the case of zero Berry phase, the trajectory of the two MSs individually form closed loops such that $\gamma_{0}^{(d)}=-\Omega_{1}/2-\Omega_{2}/2=0$, which implies that $\Omega_{1}=-\Omega_{2}$. Now the crucial differences lay in the case of Berry phase being $\pi$. In case of quadrupolar subspace, this Berry phase is induced by MS trajectories which lead to their exchange hence the two MS collectively forming a closed loop, while in the case of real subspace, in general, each MS trajectory individually form a closed loop such that the sum $\gamma_{0}^{(d)}=-\Omega_{1}/2-\Omega_{2}/2$ is constrained to $\pm\pi$. Hence it leaves us with a possibility that a pair of closed loop trajectories traced out by the MSs subjected to the $\gamma_{0}^{(d)}=\pm\pi$ in real subspace will evolve into a single exchange trajectory in the quadrupolar subspace under the action of a continuous unitary transformation.
We construct unitary transformations parameterized by a parameter $\alpha$ which can be continuously varied such that we start from the quadrupolar space and reach the real space. The unitary matrix can be parameterized as follows,
\begin{equation}\label{Parametrization}
    U(\alpha) = e^{i \theta_1 \alpha} P_1 + e^{i\theta_2 \alpha} P_2 + e^{i\theta_3 \alpha} P_3
\end{equation}
where $e^{i\theta_i}$s are the eigenvalues of the unitary in Eq.~\ref{GlobalUnitary} and $P_i$s are the projectors into corresponding eigenspaces.
\begin{align}
    P_1 =& \begin{pmatrix}
        \frac{3+\sqrt{3}}{6} & 0 & \frac{1 - i}{2\sqrt{3}}\\
        0 & 0 & 0\\
        \frac{1 + i}{2\sqrt{3}} & 0 & \frac{3-\sqrt{3}}{6}
    \end{pmatrix},
    P_2 = \begin{pmatrix}
        \frac{3-\sqrt{3}}{6} & 0 & -\frac{1 - i}{2\sqrt{3}}\\
        0 & 0 & 0\\
        -\frac{1 + i}{2\sqrt{3}} & 0 & \frac{3+\sqrt{3}}{6}
    \end{pmatrix},\nonumber\\
    &\phantom{blankspace} P_3 = \begin{pmatrix}
        0 & 0 & 0\\
        0 & 1 & 0\\
        0 & 0 & 0
    \end{pmatrix}
\end{align}
The unitary varies from identity matrix to the unitary given in Eq.\ref{GlobalUnitary} as $\alpha$ changes from 0 to 1. Applying this parameterized unitary on the MS trajectories of quadrupolar states, we shows the gradual evolution of the loops and as seen in Fig.~\ref{Parametrizedfig} and observe that the final transition into two loops occurs at $\alpha =1$. 
\section{Quadrupolar subspace dynamics} \label{sectionDynamics}
We now address the question of dynamics within the quadrupolar subspace beginning with the question that if we start with a quadrupolar state what Hamiltonians will keep the system in such a state throughout its dynamics.
The answer is completely straightforward with the help of the mapping we have identified between quadrupolar and real subspaces.
The analogous question in the real subspace becomes - ``if we start with a real state what Hamiltonians will keep it real?''
From the time-dependent Schr\"odinger equation it is easy to see that all purely imaginary Hamiltonians have this property.
For such Hamiltonians the time-dependent Schr\"odinger equation becomes a set of coupled linear ``real'' differential equations and if the initial state is real, then the subsequent state of the system remains real always.
With the help of the mapping given in Eq.~\ref{GlobalUnitary}, the above answer gets translated to all spin Hamiltonians, i.e., Hamiltonians with no quadrupolar term.
Physically this result is slightly non-trivial: while the eigenstates of a quadrupolar Hamiltonians can be chosen to be quadrupolar and would not evolve non-trivially in time, a general quadrupolar state which is not an eigenstate will become non-quadrupolar upon dynamics with a general quadrupolar Hamiltonian.
In other words, the quadrupolar Hamiltonians do not preserve the quadrupolar nature of states during time-evolution.
On the other hand with any pure spin Hamiltonian, all quadrupolar states will remain quadrupolar throughout their dynamics.
This result is consistent with the recent findings that pure spin Hamiltonians generate rigid rotation of the Majorana sphere~\cite{Dogra_2018}.
As a result a pair of antipodal stars will remain so throughout the dynamics with such Hamiltonians.

We next consider the issue of geometric phase accumulated during cyclic evolution of quadrupolar states due to application of static pure spin Hamiltonains discussed above in the spirit of Aharonov-Anandam type scenario~\cite{Aharonov1987}.

For a purely spin Hamiltonian, a quadrupolar state remains quadrupolar during dynamics.
If the Hamiltonian generates a periodic orbit, the total phase accumulated can only be $0$ or $\pi$.
Also, by definition the mean energy of pure spin Hamiltonians for any state in quadrupolar subspace is zero (Eq.\ref{QuadrupolarStates}).
Hence there is no dynamical phase contribution and the total phase is infact the geometric phase.
Again as before it is easier to answer this question for the real subspace and then translate the result to the quadrupolar one using the global change of basis.
The three spin operators (see appendix \ref{spin_operators}) upon transformation into real space using Eq.\ref{GlobalUnitary} become
\begin{equation}\label{SpinOperatorsReal}
S'_x = \left[\begin{matrix}0 & i & 0\\- i & 0 & 0\\0 & 0 & 0\end{matrix}\right],
S'_y = \left[\begin{matrix}0 & 0 & 0\\0 & 0 & - i\\0 & i & 0\end{matrix}\right],
S'_z = \left[\begin{matrix}0 & 0 & - i\\0 & 0 & 0\\i & 0 & 0\end{matrix}\right]~,
\end{equation}
and we write the most general purely imaginary Hamiltonian as
\begin{equation}\label{spin-H}
H = -a S'_x + b S'_y + c S'_z = \left[\begin{matrix}0 & - i a & - i c\\i a & 0 & - i b\\i c & i b & 0\end{matrix}\right]~.
\end{equation}
The eigenvalues of this Hamiltonian are $0, \pm \omega$, where $\omega = \sqrt{a^2 + b^2 + c^2}$.
The initial real state
$|\psi(0) \rangle = \left[\begin{matrix}r_0, s_0, t_0\end{matrix}\right]^\mathrm{T}$
at time $\tau$ becomes
\begin{widetext}
\begin{equation} \label{psiTau}
|\psi(\tau) \rangle = \frac{1}{\omega^{2}} \left[\begin{matrix}a b t_{0} + b^{2} r_{0} - b c s_{0} - \omega \left(a s_{0} + c t_{0}\right) \sin{\left (\omega \tau \right )} + \left(a^{2} r_{0} - a b t_{0} + b c s_{0} + c^{2} r_{0}\right) \cos{\left (\omega \tau \right )}\\- a c t_{0} - b c r_{0} + c^{2} s_{0} + \omega \left(a r_{0} - b t_{0}\right) \sin{\left (\omega \tau \right )} + \left(a^{2} s_{0} + a c t_{0} + b^{2} s_{0} + b c r_{0}\right) \cos{\left (\omega \tau \right )}\\a^{2} t_{0} + a b r_{0} - a c s_{0} + \omega \left(b s_{0} + c r_{0}\right) \sin{\left (\omega \tau \right )} + \left(- a b r_{0} + a c s_{0} + b^{2} t_{0} + c^{2} t_{0}\right) \cos{\left (\omega \tau \right )}\end{matrix}\right]~,
\end{equation}
\end{widetext}
which remains real and oscillates with frequency $\omega$.

Any initial state comes back to itself at $T = 2\pi / \omega$, and the question of a quantized geometric phase of $\pi$ boils down to whether it becomes the exact negative of the initial state at some intermediate time.
To answer this question we measure the distance of $|\psi(\tau) \rangle$ from $-|\psi(0) \rangle$, i.e. $|| |\psi(\tau) \rangle + |\psi(0) \rangle ||^2$ with $\tau$, for a given initial condition and Hamiltonian to obtain an expression of the form $A + B \cos(\omega \tau)$, where $A, B$ depend on the Hamiltonian parameters and the initial condition.
Since the distance we are interested in is maximum at $\tau = 0$, it initially decreases and acquires its minimum value at $\tau = T/2 = \pi / \omega$.
The minimum value of this distance becomes zero when the initial condition obeys
$$a t_0 = - b r_{0} + c s_{0}.$$
When this condition is satisfied Eq.\ref{psiTau} in fact becomes an equation for a geodesic (see Eq.~4.16 in Ref.~\cite{Mukunda1993a}) passing through the initial state
\begin{equation}\label{geodesic}
|\psi(\tau) \rangle = \cos(\omega \tau) \left[\begin{matrix}r_{0}\\s_{0}\\t_{0}\end{matrix}\right]
+ \frac{\sin(\omega \tau)}{a \omega} \left[\begin{matrix}- a^{2} s_{0} + b c r_{0} - c^{2} s_{0}\\a^{2} r_{0} + b^{2} r_{0} - b c s_{0}\\a \left(b s_{0} + c r_{0}\right)\end{matrix}\right].
\end{equation}
The normalization of wavefunction restricts the set of initial conditions to a two-dimensional surface that can be conveniently considered to be the surface of a unit sphere centered at origin.
The above condition of obtaining loops with quantized geometric phase of $\pi$ is in fact an equation of a plane passing through origin hence identifying a great circle on this unit sphere.  
For all initial conditions the Hamiltonian generates a closed trajectory on this sphere but only those initial conditions which lead to a great circle as the  closed trajectory are of interest and are given by Eq.~\ref{geodesic}.
We can easily translate this result to the  quadrupolar subspace by using the unitary transform in Eq.~\ref{GlobalUnitary}.
\section{Conclusion} \label{sectionSummary}
To summarize we have shown that the quadrupolar subspace is topological and there is a convenient geometrical way of visualizing this using MSR.
We find a global unitary transformation that relates the quadrupolar subspace to the real subspace, which is yet another way to reach the same conclusion. Using this global unitary transformation we identify the anti-unitary symmetry which ensures the topological nature of the quadrupolar subspace. 
We show that in MSR, some well-known topological properties becomes very easy to visualize geometrically, such as all close looped trajectories of eigenstates in the parameter space of a three-level system can not be simultaneously topological. It can have either 0 (trivial) or 2 (non-trivial) topological eigenstates.
We have shown that, though the quadrupolar and the real subspace are related by global unitary transformation but the trajectories of MSs on the Bloch sphere connected by this transformation can be quite distinct as depicted in Fig.~\ref{Parametrizedfig}. 
Finally we demonstrate that any static pure spin Hamiltonain acting on a quadrupolar state results in time evolution which stays restricted to the quadrupolar subspace and closed loop time evolution results in either zero or $\pi$ geometric phase.

\section*{Acknowledgments}
R.S. thanks Sthitadhi Roy, Krishanu Roychowdhury and Subhro Bhattacharjee for many helpful discussions. S.D. would like to thank  Michael V. Berry for stimulating discussions and for pointing out Ref.~\cite{Robbins_1994}. R.S. thanks the Science and Engineering Research Board (SERB), India for funding via the Ramanujan Fellowship (SB/S2/RJN-034/2016). S.D. would like to acknowledge the MATRICS
grant (MTR/ 2019/001 043) from the Science and
Engineering Research Board (SERB) for funding.    
\appendix
\section{Antipodal MSs for quadrupolar states}\label{appendix_antipodal_MSs}
Here, we discuss the implications of the  global unitary map (Eq.~\ref{GlobalUnitary}) between the real subspace  and the quadrupolar subspace in terms of the MSR of the corresponding states.
Consider a most general real state $|\psi\rangle_{R}=(r,s,t)^{T}$ of a spin-1 system, where $r, s$ and $t$ are real. This real states maps on to a quadrupolar state as

\begin{equation}
 |\psi\rangle_{Q}=U^{\dagger}|\psi\rangle=\frac{1}{\sqrt{2}}\begin{pmatrix}-t-ir \\ \sqrt{2}s \\ t-ir\end{pmatrix}~.
\end{equation}
So the Majorana polynomial equation for $|\psi\rangle_{Q}$ is 
\begin{equation}
    \frac{(-t-ir)x^{2}}{2}-sx+\frac{(t-ir)}{2}=0~,
\end{equation}
whose roots are given by
\begin{equation}
    X_{1,2}=\frac{-s\pm\sqrt{s^{2}+r^{2}+t^{2}}}{r^{2}+t^{2}}(t-ir)~.
\end{equation}
Now we rewrite the roots in the form $\tan{\frac{\theta}{2}}e^{i\phi}$, where $\theta$ and $\phi$ corresponds to the polar and azimuthal coordinates of MSs on the Bloch sphere. Hence the spherical polar coordinates for the two MSs are given by
\begin{eqnarray*}
\theta_{1}=2\tan^{-1}(\frac{s+\sqrt{s^{2}+r^{2}+t^{2}}}{\sqrt{r^{2}+t^{2}}}),\ \phi_{1}=\arg(-t+ir)~,\\
\theta_{2}=2\tan^{-1}(\frac{-s+\sqrt{s^{2}+r^{2}+t^{2}}}{\sqrt{r^{2}+t^{2}}}),\ \phi_{2}=\arg(t-ir)~.
\end{eqnarray*}
From above polar coordinates of MSs, we get $\theta_{1}+\theta_{2}=\pi$ and $\phi_{2}-\phi_{1}=\pi$ i.e. the two MSs are antipodal.
Now let us consider the real state $(1/2,1/\sqrt{2},1/2)^{T}$ whose Majorana polynomial equation is
\begin{equation}
    \frac{x^{2}}{2\sqrt{2}}-\frac{x}{\sqrt{2}}+\frac{1}{2\sqrt{2}}=0
\end{equation}
Its both roots are $X=1$. So the corresponding Majorana stars are coincident.
Hence the  global unitary transform takes two coincident MSs to antipodal position, when the real state is mapped to quadrupolar state.
\section{Spin operators}\label{spin_operators}
The spin operators in the $S_z$ basis are give by,
\begin{equation}   \label{SpinOperators}
S_x = \frac{1}{\sqrt{2}} \left[\begin{matrix}0 & 1 & 0\\1 & 0 & 1\\0 & 1 & 0\end{matrix}\right],
S_y = \frac{1}{\sqrt{2}} \left[\begin{matrix}0 & - i & 0\\i & 0 & - i\\0 & i & 0\end{matrix}\right],
S_z = \left[\begin{matrix}1 & 0 & 0\\0 & 0 & 0\\0 & 0 & -1\end{matrix}\right].
\end{equation}

Real space representation of these operators are given in Eq.~\ref{SpinOperatorsReal}.

\section{MS Exchange induced Berry phase}\label{MS_Exchange}

Here we show that $\gamma_{c}^{(0)}$ evaluated for closed loop path in quadrupolar ray space, which corresponds to a Berry phase of $\pi$, can not be expressed as sum of geometric phase arising from evolution of the individual MSs. It is rather a collective Berry phase.  We start by noting that the  Berry connection corresponding to each MS is given by 

\begin{equation}\label{BerryConnection}
 \gamma_{c}^{(0)} = \text{Im}\langle\psi|\nabla\psi\rangle=\text{Im}\sum_{i}\langle \hat{u}_{i}|d\hat{u}_{i}\rangle
\end{equation}

\noindent where, $\hat{u}_{i}$ is the unit vector from the origin to the position of MS on the Bloch sphere which represents the state

\begin{equation}
 \left|\hat{u}\right>=\cos\frac{\theta}{2}\left|\uparrow\right>+e^{i\phi}\sin\frac{\theta}{2}\left|\downarrow\right>
   \label{BlochState}
\end{equation}

\noindent and its differential in parameter space of MSs is

\begin{equation}\label{DiffBloch State}
   |d\hat{u}\rangle=-\frac{1}{2}\sin\frac{\theta}{2}d\theta\left|\uparrow\right>+e^{i\phi}\left(\frac{1}{2}\cos\frac{\theta}{2}d\theta+i\sin\frac{\theta}{2}d\phi\right)\left|\downarrow\right>~,
\end{equation}
where $\theta$ and $\phi$ are the polar coordinates of the unit vector $\hat{u}$. Then from Eq.(\ref{BerryConnection}), the Berry connection is 

\begin{equation}
    \text{Im}\sum_{i}\langle \hat{u}_{i}|d\hat{u}_{i}\rangle=\sum_{i}\frac{1-\cos\theta_{i}}{2}d\phi_{i}~.
\end{equation}

The Berry phase is the line integral of Berry connection over a closed path, and if and only if each MSs completes a loop individually on the Bloch sphere, then it can be expressed as a sum of half the solid angles made by each stars.

Now for the case of exchange of MSs under closed loop evolution of quadrupolar states in the ray space, the Berry phase is given by the sum of integral of the Berry connection for each star over a open paths (not closed) given by 

\begin{equation}\label{BerryIntegral}
    \gamma_{c}^{(0)}=-\text{Im}\int_{1}^{2}\langle \hat{u}_{1}|d\hat{u}_{1}\rangle-\text{Im}\int_{2}^{1}\langle \hat{u}_{2}|d\hat{u}_{2}\rangle~.
\end{equation}
As the two starts are constrained to stay antipodal while the states evolves to exchange the MSs, hence these two integrals together forms single closed line integral such that it result in a Berry phase of $\pi$. 

\section{Correlation term in real subspace}\label{correlation_term}
In this section we show that the vanishing of term $\gamma^{(d)}_{C}$ appearing in the expression of Berry phase given in Eq.~\ref{BerryPhaseInMSR} not only holds for quadrupolar subspace but also for real subspace. It is given by, $\gamma^{(d)}=\gamma^{(d)}_{0} + \gamma^{(d)}_{C}$ 

\begin{equation*}
\gamma_{C}^{(d)}=\frac{1}{2} \oint \sum_{i=1}^{d-1} \sum_{j(>i)}^{d-1} \beta_{i j} \Omega\left(\mathrm{d} \hat{u}_{i j}\right)
\end{equation*}
Here, $\beta_{i j}$ is the correlation factor (see Eq.\ref{correlation}) and $\Omega\left(\mathrm{d} \hat{u}_{i j}\right) \equiv \hat{u}_{i} \times \hat{u}_{j} \cdot d\left(\hat{u}_{j}-\hat{u}_{i}\right) / d_{i j}$. It can be interpreted as the solid angle due to the relative motion between each pair of stars and their absolute evolution~\cite{Liu2014} (as described in subsection \ref{4A}).

In the quadrupolar space we get antipodal MSRs and hence the correlation term is trivially zero since $\hat{u}_{1} \times \hat{u}_{2}=0$. We will show now that this is also true for real subspace. This comes from the interpretation of the correlation term as the solid angle  due to relative motion of the two stars, which we prove to be zero when restricted to real space. Hence the problem reduces to showing that this solid angle is zero.
Consider an arbitrary real state given by
\begin{equation*}
    |\psi\rangle = \begin{pmatrix}r\\ s\\t \end{pmatrix}~.
\end{equation*}
Consider the corresponding Majorana polynomial equation given by 
\begin{equation*}
    \frac{r}{\sqrt{2}}x^2 - s x + \frac{t}{\sqrt{2}}=0~,
\end{equation*}
 which is a quadratic equation with real co-efficients in our case. Hence the roots are either complex conjugates pair or both real.

Next we note that the identification of the ($\theta,\phi$) coordinates  of the MSs are given by the equation

\begin{align}\label{roots}
\begin{split}
      \tan\frac{\theta_1}{2}e^{i\phi_1} = \frac{s+\sqrt{s^2 - 2rt}}{\sqrt{2} r}~,\\
    \tan\frac{\theta_2}{2}e^{i\phi_2} = \frac{s-\sqrt{s^2 - 2rt}}{\sqrt{2} r}~.
\end{split}
\end{align}
The action of complex conjugation on the roots imply,
\begin{equation}
    \theta\rightarrow\theta\quad\phi\rightarrow-\phi~.
\end{equation}
This action is exactly a reflection about  the $x$-$z$ plane on the Bloch sphere. Since real states are invariant under this action we can conclude that the two MSs corresponding to states in the real subspace are always mirror reflections of each other about the $x-z$ plane. 

Now, the possibility of the relative motion of the MSs forming finite solid angles is severely restricted. One of the possible case is when both stars remain entirely in $x$-$z$ plane throughout and the relative vector extends a $2\pi$ solid angle during the path. However, the correlation term becomes zero here since $\Omega\left(\mathrm{d} \hat{u}_{i j}\right)=\hat{u}_{i} \times \hat{u}_{j} \cdot d\left(\hat{u}_{j}-\hat{u}_{i}\right) / d_{i j}$ is zero as $\hat{u}_{i} \times \hat{u}_{j}$ points perpendicular to the $x$-$z$  plane while $d\left(\hat{u}_{j}-\hat{u}_{i}\right)$ is always in the plane. On the other hand if the MSs stay out of the the $x$-$z$, then the relative vector between the MRs are constrained by the reflection symmetry and hence is left with a one dimension degree of freedom passing through the origin leading zero solid angle. This completes our proof of $\gamma^{(d)}_{C}$  being zero for the real states.

\nocite*
\bibliography{apssamp.bib}

\end{document}